\documentstyle[11pt,Cargesepasp,twoside,epsf]{article}
\def\msol{\mbox{M}_\odot}

\def\te{T_{\rm eff}}
\def\simgr{\,\hbox{\hbox{$ > $}\kern -0.8em \lower 1.0ex\hbox{$\sim$}}\,}
\def\simle{\,\hbox{\hbox{$ < $}\kern -0.8em \lower 1.0ex\hbox{$\sim$}}\,}

\def\edcomment#1{\iffalse\marginpar{\raggedright\sl#1\/}\else\relax\fi}
\reversemarginpar

\begin{document}
\title{Pre-Main Sequence models for low-mass stars and brown dwarfs}
 \author{Isabelle Baraffe}
\affil{Ecole Normale Sup\'erieure, CRAL,69364 Lyon Cedex 07, France}
\author{Gilles Chabrier}
\affil{Ecole Normale Sup\'erieure, CRAL,69364 Lyon Cedex 07, France}
\author{France Allard}
\affil{Ecole Normale Sup\'erieure, CRAL,69364 Lyon Cedex 07, France}
\author{Peter Hauschildt}
\affil{Center for Simulational Physics, University of Georgia
Athens, GA 30602-2451}

\begin{abstract}
We present evolutionary models for low mass stars and brown dwarfs
($m \le 1.2 \msol$) based on recent improvement of the theory: equation of state, atmosphere models, ...
We concentrate on early evolutionary phases from the initial deuterium burning phase to the zero-age Main Sequence. Evolutionary models for young brown dwarfs
are also presented. We discuss the uncertainties of the present models.
We analyse the difficulties arising when comparing
models with observations for very young objects, in
particular concerning the problem of reddening.


\end{abstract}

\section{Introduction}

Within the past years, important efforts have been devoted to the observation
and the theory of very-low-mass stars (VLMS) and substellar objects (brown
dwarfs BD and giant planets GP). The main
theoretical improvements involve the  description 
of the interior of these cool and dense objects (equation
of state for dense plasmas, screening factors, etc...; see the
 review by Chabrier and 
Baraffe 2000) and the model atmosphere (molecular opacity, formation of
dust, etc...; see the review by Allard et al. 1997). A major advance
in the field is  
the development
of a new generation of consistent models based on the coupling of interior
and atmosphere models,  providing direct comparison of evolutionary models
with observations in  colour-colour and colour-magnitude diagrams (CMD). 
Several observational tests, mainly provided by relatively old objects
(age $\simgr$ 100 Myr), now assess the validity of this theory devoted
to stellar and substellar objects with masses $\le 1 \msol$. 
General agreement is found with (i) the mass - radius relationship of
observed eclipsing binaries, (ii) mass - magnitude relationships
in $VJHK$ provided by binary systems, (iii) mass - spectral type
relationships for M-dwarfs, (iv) colour - magnitude relationships
for intermediate age open clusters (Pleiades, Hyades, etc...), field
disk M- and L-dwarfs, halo stars, and globular cluster Main Sequences
 and (v) spectra of M-dwarfs.
Details and references for such confrontations of models with observations
can be found in Chabrier and Baraffe (2000). Although some discrepancies
between models and observations remain  (see \S 2), uncertainties due
to the input physics are now significantly reduced.  

Numerous surveys devoted to the search for substellar objects
 have been conducted
in young clusters with ages spanning from $\sim$ 1-10 Myrs,
providing a wealth of data for pre-Main Sequence (PMS) objects.
The reliability of the present theory for VLMS and BD
 allows now a thorough analysis
for such PMS objects. Unlike to older Main Sequence stars and BDs,
comparison between observations and models for very
young objects presents some difficulties:
(i) extinction due to the surrounding dust
modifies both the intrinsic magnitude and the colours of the object,
(ii) gravity affects both the spectrum and the evolution, (iii)
the evolution and
spectrum of
 very young objects ($t \simle$ 1 Myrs) may still be affected by the presence
of an accretion disk or circumstellar material residual from the protostellar 
stage.

This paper is devoted to PMS models for VLMS and BDs and
completes the work of Baraffe et al. (1998, BCAH98) which was essentially
devoted to the comparison of models with 
observations of older objects ($t \simgr$ 100 Myrs).
We discuss the remaining uncertainties of the models (\S 2) 
and analyse their confrontation with available observations  (\S 3).    

\section{Pre-Main Sequence models}

\subsection{Evolutionary tracks}

The present models are based on the input physics described in 
Baraffe et al. (1998, and references therein).
The robustness of these models is anchored in two areas: the microphysics
determining the equation of state (EOS) in the stellar interior, and
the outer boundary condition and  synthetic spectra
based on non-grey atmosphere models (Hauschildt,
 Allard \& Baron 1999). Figure 1 shows evolutionary tracks 
from 0.02 $\msol$ to 1 $\msol$. The stellar/substellar transition is located
at $\sim$ 0.075 $\msol$, below which objects become partially degenerate
and never reach thermal equilibrium characterising the Main Sequence. 
Evolutionary models start at the beginning of the initial deuterium burning
phase. The deuterium  burning minimum mass is $\sim$ 0.013 $\msol$ 
(Saumon et al. 1996; Chabrier \& Baraffe 2000). The D burning phase lasts less than
1 Myr for $m \simgr 0.2 \msol$, between 1 and 5 Myr for  $0.05 \simle m \simle 0.2 \msol$
and almost 20 Myr for a 0.02 $\msol$ brown dwarf. At such young ages, evolution
is characterised by a rapid contraction of the object once central D is
significantly depleted. Consequently, there is a significant variation of the surface
gravity from $\log g \, \sim \,  3$ to $\sim$ 4.5 from 1 Myr to 50 Myr,
for the masses displayed in Fig. 1. Since the present atmosphere models
(Hauschildt et al. 1999) assume the plane-parallel approximation, we
have checked its validity even for such low gravities ($\log g \, \sim \,  3$).
 A comparison of
these models with atmosphere models including effects of spherical geometry
shows that the latter are important only for surface gravities 
$\log g \, \simle \,  2$ (Hauschildt, Allard, Ferguson, et al. 1999). 
This is well below the range of gravities involved
in the evolution of the present objects. 

\begin{figure} 
\plotone{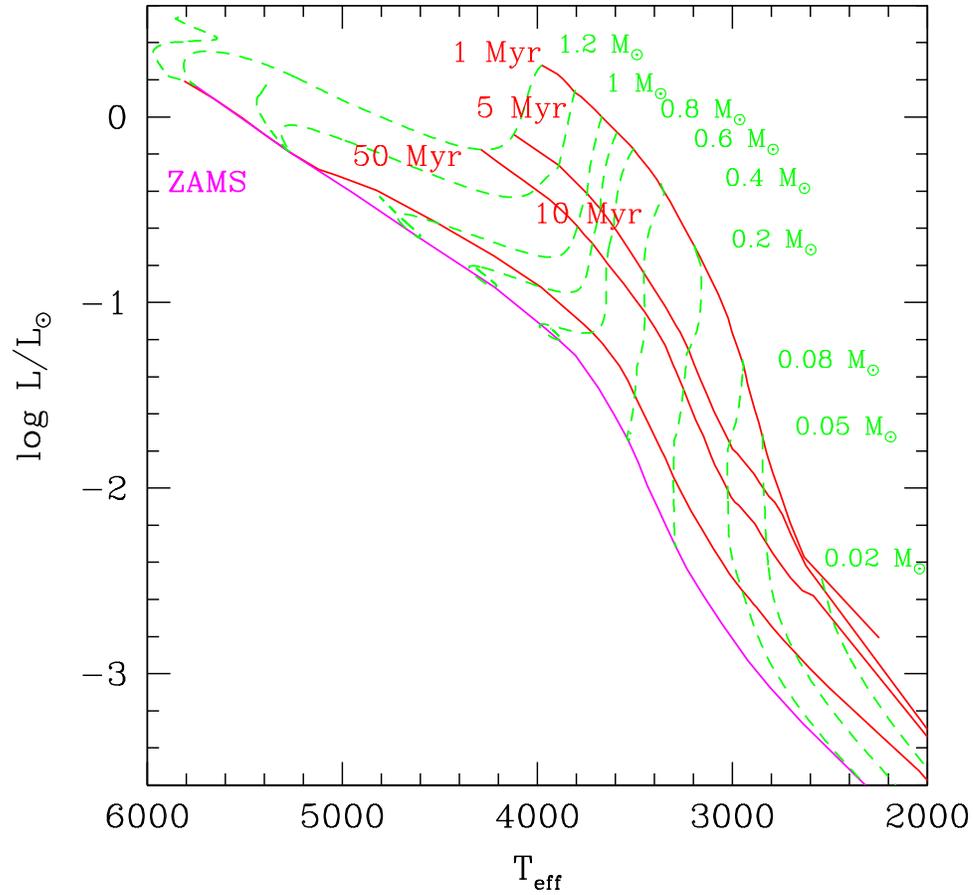}
\caption{Evolutionary tracks in the Hertzsprung-Russell diagram for masses
from 1.2 $\msol$ to 0.02 $\msol$ (dashed lines) 
and ages spanning from 1 Myr to the ZAMS
(for stars). Several Isochrones for 1, 5, 10 and 50 Myr
are indicated by solid lines from right to left. The location of the ZAMS,
for stars down to 0.075 $\msol$, is also indicated (left solid line). }
\end{figure}

\subsection{Improvement and uncertainties}

As already mentioned, one of the main improvement in the modeling of
VLMS and BDS is the use of outer boundary conditions based on
realistic non-grey atmosphere models. As demonstrated by
Chabrier \& Baraffe (1997; and references therein) the use
of radiative $T(\tau)$ relationships and/or grey atmosphere models
is unvalid  when molecules form near the photosphere, below $\te \sim 4000K$.
Outer boundary conditions based on the latter approximations
yield hotter models for a given mass, as illustrated in Fig. 2a. 
Interestingly enough, masses up to 1 $\msol$ are affected by the choice
of the outer boundary condition, since at young ages  evolution proceeds at
significantly lower $\te$ than on the MS for $m > 0.5 \msol$ (see Fig. 1).  
The use of an inappropriate outer boundary condition,
 such as the Eddington
approximation, yields an overestimation of $\te$ for a given $m$
up to 300 K (see Fig. 2a).  

One of the main uncertainty for masses $m \ge 0.7 \msol$ is
due  to convection. These stars show relatively extended
superadiabatic outer layers, which are extremely sensitive to the treatment of
convection. In the framework of the mixing length formalism, this translates
into a sensitivity to the mixing length $l_{\rm mix} \, \propto \, H_{\rm P}$,
with $H_{\rm P}$ the pressure scaleheight. Figure 2b illustrates
the effect of a variation of $l_{\rm mix}$ on PMS tracks.
 For the present models, $l_{\rm mix}$ = 1.9 $H_{\rm P}$
is the value required to fit the Sun at its present age. 
Recently, Ludwig, Freytag \& Steffen (1999)
calibrated the mixing length parameter $\alpha = l_{\rm mix}/H_{\rm P}$
with 
2D hydrodynamical models performed 
in the parameter space 4300 K $\le \te \le$ 7100 K and gravities
2.54 $\le \log g \le$ 4.74. They found a moderate variation of the
mixing length parameter around typically 1.5. Fig. 2b shows
that a variation of $\alpha$ from 1 to 1.9 yields an increase
of $\te$ up to 500K for the highest masses during their
PMS evolution.
For masses $m \simle 0.7 \msol$, the size of the superadiabatic
layers reduces and the transition from the  convective to the
radiative outer layers
is characterised by an abrupt transition from a fully adiabatic to
a radiative structure. The sensitivity of the models to $l_{\rm mix}$ is
thus small, in the framework of the mixing length formalism. 
To confirm this low sensitivity, however, multi-dimensional simulations
as done by Ludwig et al. (1999) must clearly be extended below
 $\te \, \le$ 4300 K.

\begin{figure} 
\plotone{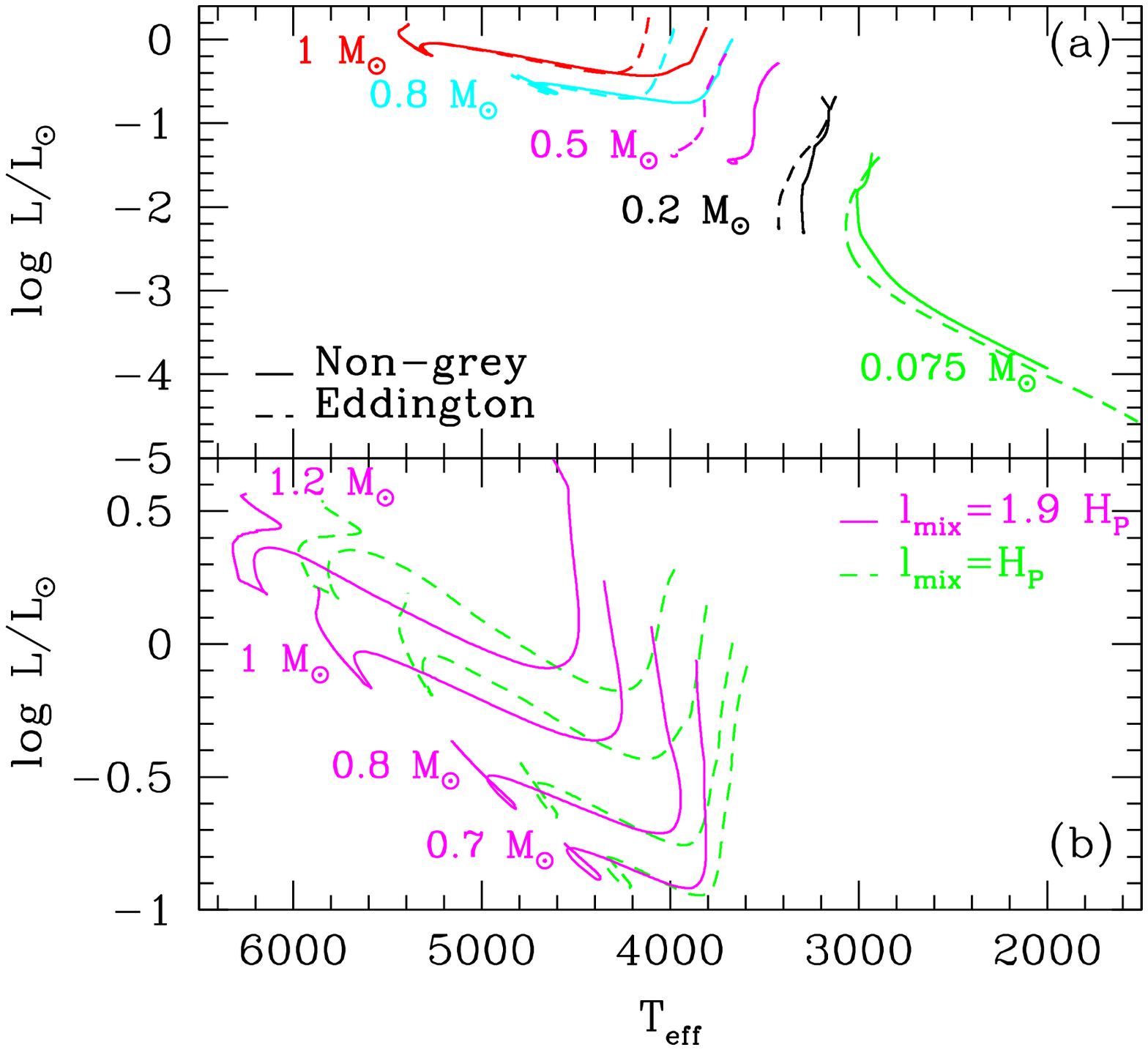}
\caption{{\bf (a)} Effect of the outer boundary condition on evolutionary
tracks. {\bf (b)} Effect of a variation of the mixing length $l_{\rm mix}$.}
\end{figure}

Finally, although huge improvements were made in the field of molecular
opacities (see Allard et al. 1997), considerably reducing  the uncertainties
of non-grey atmosphere models and synthetic spectra, some shortcomings
still remain. As mentioned in BCAH98, the present models still show a 
$\sim$ 0.5 mag
discrepancy with observations in optical $VRI$
 colour - magnitude diagrams. This
problem was partly identified as a shortcoming in the TiO linelist, one of
the main absorber in the optical. A new TiO linelist recently
computed by Schwenke (1998) improves indeed the situation (see Fig. 3a), 
although some discrepancies with observations in $(V-I)$ (see Fig. 3a)
and $(R-I)$ colours
still remain (see Chabrier et al. 2000). The effect of the new TiO line list
 is illustrated in Fig. 3a,
where  disk field objects of Monet et al. (1992) and Dahn et al. (1995) are
 also displayed. Although the new TiO linelist significantly reduces
 the discrepancy with observations (cf. Fig. 3a), the fit
of the models to the data is not perfect yet for objects
fainter than $M_{\rm V}\sim 10$. Moreover, as illustrated in Fig. 3b for near-IR
colours, the use of this new TiO linelist affects the spectrum in the near-IR,
and worsens the excellent agreement 
with observational data from the Pleiades 
in $(I-K) - M_{\rm I}$ obtained previously
with the BCAH98 models. Another uncertainty appears in the water molecular
linelist. BCAH98 uses the list of Miller et al. (1994) which is known
to be incomplete for high energy transitions. Although a  more complete linelist
was recently computed by Partridge \& Schwenke (1997), it still has
shortcomings and yields large discrepancies with photometric observations
in the near-IR above $\te \simgr 2300K$, as illustrated in Fig. 3b (dash-dotted line).  

Given the problems  with the current linelists,
improvement of the present evolutionary models have to await the computation 
of more reliable H$_2$O, and to a lesser extend TiO, linelists. Although
the different molecular linelists mentioned above affect significantly
the spectra and colours, they hardly affect the deeper
atmospheric layers, and therefore the outer boundary condition
to the interior. Thus, their effect on evolutionary
models in terms of $\te$ and $L$ is small. The use of
 the new TiO and/or water linelist does not affect  evolutionary models 
by more than 100 K in
$\te$ and 10\% in $L$  at a given age. This illustrates the remaining
(small) uncertainty of the models due to molecular opacities
from the evolution viewpoint. Below
$\te \simle 2300K$, grain formation starts to affect both the spectrum
and the evolution, and must be taken into account (cf. Chabrier et al. 2000).
This is out the scope of the present paper, which is based 
on dust-free models. Only substellar objects with $m \, \simle \, 0.01 \msol$
and older than 1 Myr are affected by dust formation.

\begin{figure} 
\plotone{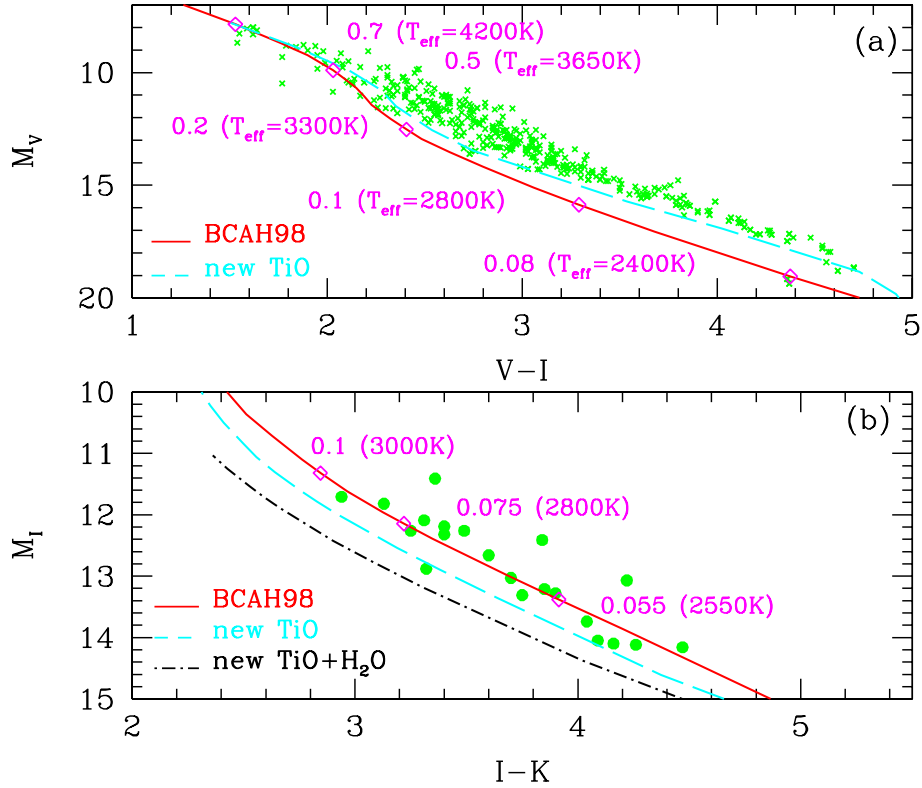}
\caption{
{\bf (a)} Effect of the TiO line list on models in a 
$(V-I) - M_{\rm V}$
diagram. The solid line corresponds to an isochrone of 1 Gyr based
on the present models (BCAH98). The dashed line correspond to models
computed with the new TiO
line list, for the same age. Observations (crosses) are disk field stars
from Monet et al. (1992)
and Dahn et al. (1995). Masses (in $\msol$) and $\te$ are indicated and correspond
 to the open
diamonds on the solid line.   
 {\bf (b)} Effect of TiO and H$_2$O line lists on models in
a $(I-K) - M_{\rm I}$ diagram. The solid and dashed lines are the same models
as in (a) for an age of 120 Myr. The dash-dotted line corresponds to models with  new 
TiO and H$_2$O linelists, for 120 Myr. Observational data (full circles)
belong to the Pleiades, corresponding to an age of 120 Myr (Mart\'in et
al. 2000).
}
\end{figure}

\section{Comparison with observations}

The previous section describes uncertainties inherent to the input
physics of VLMS and BDs models. For  PMS
models specifically, another important source of 
uncertainty comes from the choice of initial models.
The present models are based on extremely simplified assumptions, 
starting shortly  before the deuterium burning phase in
an initially fully convective configuration, and neglecting the protostellar
accretion
phase. Theses assumptions are standard (see also Siess et al. 2000; Siess 2000, this conference). Some attempts exist
 to use more sophisticated initial conditions 
based on protostar models 
(see Palla \& Stahler 1999; and references therein). Given however
the complexity of star formation processes and 
protostellar collapse calculations (see Masunaga, Miyama \& Inutsuka 1998; Masunaga \& Inutsuka 2000; and references therein; Inutsuka, this conference;
Wuchterl, this conference), many problems are  unsolved to date
and the link between the dynamical phase of collapse and the quasi-static
phase of evolution is still very obscure. 

Testing different initial conditions, we find out 
that after a few Myr they become inconsequential and
models converge toward the same track. However, these tests are still
based on relatively simple assumptions. The comparison of pre-MS tracks 
with observations of
very young objects can improve our understanding of the protostellar
collapse phase 
and can tell us at which stage initial conditions become important.   

Because of the problem of large reddening in young clusters,
direct comparison of observations with models in colour - magnitude diagrams
are extremely uncertain. There are only a few exceptions, such as
$\sigma$ Orionis which exhibits low extinction. Recently, B\'ejar et al. (1999)
and Zapatero et al. (1999) obtained optical and near-IR photometry
for low mass objects in this cluster. 
The data are well reproduced by a 5 Myr isochrone based on the BCAH98 models
 in a $(I-J) - M_{\rm I}$ CMD (see Fig. 1 of Zapatero et al. 1999;
 Zapatero, this
conference),   down to 0.015 $\msol$. Such observations
 are extremely important, 
since they provide the best opportunity
to determine the Initial Mass Function (IMF) down to the substellar regime
(see B\'ejar et al. 2000).  Indeed, for such young clusters, no significant 
dynamical evolution is expected and their mass function should be close to
the true IMF. 

Young multiple systems provide also excellent tests for
PMS models at young ages, because of the expected coevality of
their different components. In addition, another strong constraint 
is supplied by the estimate of
dynamical masses  based on the orbital motion of 
circumstellar/circumbinary disks (Simon, Dutrey\&  Guilloteau 2000). 
One of the best example is provided by
the quadruple system GG TAU (White et al 1999),
with components covering the whole mass-range of VLMS and BDs from 1 $\msol$
to $\sim$ 0.02 $\msol$. Orbital velocity measurements of the circumbinary
disk surrounding the two most massive components imply a constraint
on their combined stellar mass (Dutrey, Guilloteau \& Simon 1994;
 Guilloteau, Dutrey \& Simon 1999). This mass constraint and the hypothesis
of coevality provide a stringent test for PMS models. 
Models based on non-grey atmospheres (BCAH98) are
 the only ones consistent with these observations
(for details see White et al 1999;  Luhman 1999).
 
Two major difficulties remain however when comparing the  models 
with such data: 
(i) the spectral type classification  and (ii) its transformation
in $\te$ based on a $\te$ - scale. Young objects are expected to
show spectral
features between that of giants and dwarfs, and a better representation
of their spectral properties may require a new classification 
more appropriate to these intermediate surface
gravities. The transformation of the inferred spectral type in $\te$
is even more difficult, because of the lack of reliable 
$\te$ - scales for such young T Tauri like objects. Significant
efforts were devoted  within the past years to the elaboration
of improved $\te$ - scales for M-dwarfs (Leggett et al. 1996) 
and M-giants (Perrin et al. 1998; van Belle et al. 1999). However, work
remains to be done for T Tauri like objects. 
Recently, Luhman (1999) defined a $\te$ - scale intermediate between giants
and dwarfs and based on the isochrone of the BCAH98 models which
passes through the 4 components of GG Tau. Interestingly enough, applying
this $\te$ - scale to young clusters such as IC348 (Luhman 1999) and
star forming regions (Chamaeleon I, Comer\'on, Neuh\"auser \& Kaas 2000),
the cluster members show a small scatter in age and no apparent
trend of a correlation between age and mass. As mentioned by Comer\'on
et al. (2000), this suggests in Chamaeleon I an almost
 coeval population which formed
within less than 1 Myr. Confirmation of this property in other young
clusters is urgently required to improve our understanding of formation
process timescale  in young stellar associations.

\section{Conclusion}

The good agreement of models based on improved physics with observations
for relatively old (t $\simgr$ 100 Myr) VLMS and BDs now yields good
confidence in the theory of these objects. Such evolutionary models
can now be confronted to the complex realm of very young objects,
thus providing important informations on star formation processes and 
initial conditions for PMS models. 
Although based on extremely simple initial conditions (no accretion phase,
no account of protostellar collapse phase and timescale),
these models are the most consistent with present observations of
very young objects
(estimate
of dynamical masses, tests of coevality in multiple systems, CMD, etc..).
Such consistency must be confirmed with more observational tests and
the elaboration of a reliable $\te$-scale, in order to guide 
protostar collapse models, which at some age must converge toward
the PMS tracks.

\medskip
{\it Note: }
Tracks and isochrones from 0.02 $\msol$ to 1.2 $\msol$ and $t \, \ge$ 1 Myr
are available by anonymous ftp:
\par
\hskip 1cm ftp ftp.ens-lyon.fr \par
\hskip 1cm username: anonymous \par
\hskip 1cm ftp $>$ cd /pub/users/CRAL/ibaraffe \par
\hskip 1cm ftp $>$ get README \par
\hskip 1cm ftp $>$ get BCAH98\_models.* \par
\hskip 1cm ftp $>$ get BCAH98\_iso.* \par
\hskip 1cm ftp $>$ quit


\bigskip

\begin{references}

\reference Allard, F., Hauschildt, P. H., Alexander, D. R., \& Starrfield, S.
 1997, \araa, 35, 137
\reference Allard F., Hauschildt P.H., Schwenke, D., 2000, \apj, submitted
\reference Baraffe I., Chabrier G., Allard F., Hauschildt P.H. 1998, \aap, 337, 403 (BCAH98)
\reference B\'ejar, V.J.S., Zapatero Osorio M.R., Rebolo R. 1999, \apjl, 521, 671
\reference B\'ejar, V.J.S., Mart\'in, E.L., Zapatero Osorio, M.R., Rebolo, R.,
Barrado y Navascu\'es, D., Bailer-Jones, C.A.L., Mundt, R., Baraffe, I., Chabrier, G., Allard, F. 2000, Science, submitted
\reference Chabrier, G., Baraffe, I. 1997, \aap, 327, 1039
\reference Chabrier, G., Baraffe, I. 2000, \araa, in press, astro-ph/0006383
\reference Chabrier, G., Baraffe, I., Allard, F., Hauschildt, P.H. 2000, \apj, in press, astro-ph/0005557
\reference Comer\'on, F., Neuh\"auser, R., \& Kaas, A.A. 2000, \aap, submitted
\reference Dahn C.C., Liebert J., Harris H.C., Guetter H.H., 1995, in The bottom
of the main sequence and below, ed. C. Tinney, Springer Verlag, p.239
\reference Dutrey, A., Guilloteau, S., \& Simon, M. 1994, \aap, 286, 149
\reference Guilloteau, S., Dutrey, A.,  \& Simon, M. 1999, \aap, 348, 570
\reference Hauschildt P.H., Allard F., Baron E. 1999, \apj, 512, 377
\reference Hauschildt P.H., Allard F., Ferguson, J., Baron, E., Alexander, D.R.
1999, \apj, 525, 871
\reference Leggett, S.K., Allard, F., Berriman, G., Dahn, C.C.,  \& Hauschildt, P.H. 1996, \apjs, 104, 117
\reference Ludwig, H.G., Freytag, B., \& Steffen, M. 1999, \aap, 346, 111
\reference Luhman KL. 1999,  \apj,  525, 466
\reference Mart\'in, E.L., Brandner, W., Bouvier, J., Luhman, K., Stauffer, J.,
\& Basri, G. 2000, \apj, in press 
\reference Masunaga, H., Miyama, S. M., \& Inutsuka, S-I. 1998, \apj, 495, 346
\reference Masunaga, H., \& Inutsuka, S-I. 2000, \apj, 531, 350
\reference Miller S., Tennyson J., Jones H.R.A, Longmor, A.J., 1994,
in Molecules in the Stellar Environment, ed. U.G Jorgensen, Lecture Notes
in Physics
\reference Monet, D. G., Dahn C. C., Vrba F. J., Harris H. C., Pier J. R.,
 Luginbuhl C. B., Ables H., D. 1992, \aj, 103, 638
\reference Palla, F., \& Stahler, S.W. 1999, \apj, 525, 772
\reference Partridge, H., and Schwenke, D.W., 1997, {\it J. Chem. Phys.}, 106, 4618
\reference Perrin, G., Coud\'e du Foresto, V., Rigway, S.T., Mariotti, J.-M., 
Traub, W.A., Carleton, N.P., \& Lacasse, M.G. 1998, \aap, 331, 619
\reference Saumon D., Hubbard W.B., Burrows A., Guillot T., Lunine J.I.,
Chabrier, G. 1996, \apj, 460, 993
\reference Schwenke, D.W., 1998, {\it Chemistry and Physics of Molecules and Grains in Space}, Faraday Discussion 109, 321
\reference Siess, L., Dufour, E., \& Forestini, M. 2000, \aap, in press, astro-ph/0003477
\reference Simon, M., Dutrey, A., \& Guilloteau, S. 2000, \apj, submitted
\reference van Belle, G.T., et al. 1999, \apj, 117, 521
\reference White RJ, Ghez AM, Reid IN, Schultz G. 1999, \apj, 520, 811
\reference Zapatero Osorio, M.R., B\'ejar V.J.S., Rebolo, R., Mart\'in, E.L., Basri, G. 1999, \apjl, 524, 115
\end{references}
\end{document}